\documentstyle[12pt,aasms4]{article}
\received{----------- }
\accepted{----------- }
\journalid{337}{ }
\articleid{11}{ }
\slugcomment{Submitted to ApJ.}
\lefthead{Christine C. Dantas, Reinaldo R. de Carvalho, Hugo V. Capelato \& Alain Mazure}
\righthead{Evidence of Substructure in the Cluster of Galaxies A3558}

\begin{document}

\title{EVIDENCE OF SUBSTRUCTURE IN\\
       THE CLUSTER OF GALAXIES A3558}

\author{Christine C. Dantas\altaffilmark{1}}
\author{Reinaldo R. de Carvalho\altaffilmark{2,3}}
\author{Hugo V. Capelato\altaffilmark{1}}
\author{Alain Mazure\altaffilmark{4}}

\altaffiltext{1}{Divis\~ao de Astrof\'{\i}sica, INPE/MCT, CP 515, S. J. dos Campos, SP 12201-970, Brazil.} 
\altaffiltext{2}{Present address: Caltech - Astronomy Dept., Pasadena, CA 91125, USA.} 
\altaffiltext{3}{On leave of absence from Observat\'orio Nacional/CNPq, DAF,
Brazil. }
\altaffiltext{4}{IGRAP/LAS Traverse du Siphon 13012, Marseille,France.}

\begin{abstract}

We investigate the dynamical properties of the cluster of galaxies
A3558 (Shapley 8). Studying a region of one square degree ($\sim$ 3
Mpc$^2$) centered on the cluster cD galaxy, we have obtained a
statistically complete photometric catalog with positions and
magnitudes of 1421 galaxies (down to a limiting magnitude of $B \sim
21$).  This catalog has been matched to the recent velocity data
obtained by Mazure et al. (1997) and from the literature, yielding a radial
velocity catalog containing 322 galaxies. We analyse the resulting
catalog in search of substructure, using different
statistical techniques. This analysis shows that the position/velocity
space distribution of galaxies shows significant substructure. 
 
A central bimodal core detected previously in a preliminary study
(\cite{dan94}) and by an analysis of the X-ray emission map
(\cite{sle94}) is confirmed using the Adaptive Kernel Technique and
Wavelet Analysis. We show that this central bimodal subtructure is
nevertheless composed of a projected feature, kinematically unrelated
to the cluster,  plus a group of galaxies probably in its initial
merging phase into a relaxed core.  The cD velocity offset with respect
to the average cluster redshift, reported earlier by several authors,
is completely eliminated as a result of our dynamical analysis. The
untangling of the relaxed core component also allows a better, more
reliable determination of the central velocity dispersion, which in
turn eliminates the ``$\beta$-problem'' for A3558. The cluster also
shows a ``preferential'' distribution of subclumps coinciding with the
direction of the major axis position angle of the cD galaxy and of the
central X-ray emission ellipsoidal distribution, in agreement with the
anisotropic merger scenario described by West (1994a).

\end{abstract}


\section{Introduction}

Clusters of galaxies stand as the ideal probes to test the presently
accepted hierarchical scenario of structure formation. According to
this picture, if $\Omega \sim$ 1, rich clusters are still being formed
by mergers of individual groups of galaxies from the field
(\cite{ric92}). Recent optical and X-ray observations seem to be in
agreement with that hypothesis, since substructure is found in a large
fraction of clusters ($\sim$40$-$75\%, see \cite{bir94}, \cite{wes94b}
and references therein).  Recent simulations (e.g. \cite{kat93}) based
on the hierarchical scenario are also beginning to show features
with properties that can be associated to substructures found in real
clusters.

Apart from the cosmological point of view, the complexity of the
multicomponent environment (gas, galaxies and dark matter) which
constitutes a cluster of galaxies can provide important clues to the
formation and evolution of individual galaxies.  It is specially
important to investigate these structural and evolutionary trends
within substructures, where galaxy alterations (morphological type,
size, and mass distribution) within the denser substructure environment
might be more clearly untangled (e.g. \cite{whi90}).

There are several statistical techniques now available to quantify
substructures in clusters. One of the first systematic works was
carried out by Geller \& Beers (1982).  Using only 2-D projected data,
they found that $\sim$40\% of the clusters in their sample show
statistically significant substructures. Later, Dressler \& Shectman
(1988) developed a statistic based on both position and velocity
information, and found similar results when applying their test to a
sample of 15 clusters. However, in a subsequent study, West \& Bothun
(1990) reached opposite conclusions.  Bird (1994) have analysed the
several proposed methods and statistics in detail and concluded that
$\sim$50\% of the studied sample (33 clusters) contains substructures.
More recently, Kriessler \& Beers (1997) constructed new 2-D contour
maps using Dressler's (1980) sample (the same sample analysed
previously by \cite{gel82}). By the application of the Adaptive Kernel
technique (\cite{sil86}), they found that 57 \% of clusters in their
sample presented significant substructure.

The goal of this work is to investigate the structural and dynamical
properties of the individual cluster of galaxies A3558 (Shapley 8), in
terms of substructure analysis. This cluster has been an object of
extensive optical and X-ray observations in the past few years (e.g.
\cite{met87}, \cite {tea90}, \cite{bar93}, \cite {bar96}, \cite{mar96},
\cite{dan94}, \cite{dan96}).  This cluster, classified as richness
class 4 in the ACO catalog (\cite{abe89}) and dominated by a central
giant cD galaxy, has attracted much interest, first, due to its
location - at the core of the Shapley Concentration, a supercluster
composed of 25 clusters of galaxies and rich in X-ray emitting clusters
(e.g. \cite{ray91}, \cite{zuc93}), and second, because of its dynamical
complexity due to the presence of substructure (e.g. \cite{bar96}).

Some authors have reported that the central cD galaxy of
A3558 has a peculiar velocity relative to the cluster average velocity
(\cite{geh91}).  This ``offset'', also found in several cD clusters,
was shown by Bird (1994) to be strongly correlated to the presence of
substructure in the cluster. Using the KMM method (e.g.  \cite{ash95}),
in order to assign galaxies to their most probable substructure, she
concluded that the velocity offset remained only in 2 out of 25
analysed clusters (including A3558).  Later on, Bardelli et al. (1996)
partitioned his A3558 velocity sample into two sub-samples and verified
that the cD velocity was consistent with the average velocity of one of
the subsamples, but the detailed relation of the original offset to the
nature of the dynamical status of the central core was never cleared.

Also, one more component of uncertainty in A3558 is its apparently
``extreme'' $\beta$-discrepancy, as described by Bardelli et al., which
could not be accounted for using the proposed correction of Bahcall \&
Lubin (1993).  The $\beta$-problem,  first noted by Gorenstein et al.
(1978) and Jones \& Forman (1984), can be stated in its simple version
(see details in e.g.  \cite{ger94}) as an incompatibility between the
observed X-ray slope of the hot intracluster gas and the theoretical
temperature ratio of galaxies to gas. This problem has often been
referred to in the literature, and some solutions have been proposed
(e.g. \cite{lub9?}, \cite{ger94}).

In this paper, we resolve the cD offset and the $\beta$-problem in
A3558, using a careful and detailed substructure analysis of a
statistically complete velocity sample of galaxies.  We present the
data reduction and methodology in Section 2.  In Section 3, we briefly
review the current available statistical techniques to look for
substructure and apply them to A3558 in Section 4.  In Section 5, we
discuss the results and a comparison of all the methods and their
significance.  We use $q_0$ = 0.1 and $H_0$ = 75 km s$^{-1}$ Mpc$^{-1}$
throughout.

\section {Data Reduction and Methodology}

\subsection {Data}

The Abell cluster A3558 (Shapley 8) has been extensively observed in the
last five years (e.g. \cite{bar93}). It is at a distance of
$\sim$ 14400 km s$^{-1}$ and is located at the core of the Shapley
Concentration, a supercluster composed of 25 clusters of galaxies.
This cluster is classified as richness class 4 in the ACO catalog
(\cite{abe89}), and is dominated by a central giant cD galaxy. The
cluster seems to be relatively rich in elliptical/lenticular galaxies
(\cite{met94}). All this is suggestive of a regular and well relaxed
rich cluster of galaxies.  We explore this hypothesis by analysing both
the projected distribution of galaxies as well as its velocity field.
Extensive spectroscopic surveys of clusters of galaxies (e.g. the ESO
surveys, ENACS, the Canadian CNOC survey)
have been carried out recently. However, a well defined magnitude
limited catalog is of fundamental importance for a follow-up
spectroscopic study, so that the investigation of the physical
properties of clusters becomes meaningful.

We have obtained two digitized images of A3558:  one with resolution of
1.67 arcsec/pixel and the other with 1 arcsec/pixel (\cite{las90}),
through scans of the deep IIIa-J plates available at
STScI\footnote{Photometry obtained using the Guide Stars Selection
System Astrometric Support Program developed at the Space Telescope
Science Institute (STScI is operated by the Association of Universities
for Research in Astronomy, Inc., for NASA).}.  Measurement of
photometric attributes are improved by the use of high resolution
images, but the noise contribution proportionally increases in the
scanning process.  For this reason, we tested the confidence limits of
the final photometric catalog by analysing both images of A3558
(defined by a field of 1$^{\circ}\times~$1$^{\circ}$, centered on the
cD galaxy).

\subsection {Detection and Classification With FOCAS}

The data reduction was performed using the FOCAS package
(\cite{val82}).  Several tests were carried out for both images and for
a simulated cluster image in order to understand how the output catalog
responded to variations in the object detection parameters
(\cite{dan96}). These tests show that the 1 arcsec/pixel resolution
image provided a higher quality information, being a compromise between
noiser measurement and better sampling.  We defined a minimum detection
area of 65 px$^2$, corresponding to the area of a circle with radius
equal to $\sim$3 times the typical seeing disc at Siding Spring.
Detection was performed at a threshold level of 3 times the sky rms.
This corresponds on average to 7 \% of the local sky value.  Besides
the detection procedure, an important step in building a reliable
catalog is to set the ``Point Spread Function'' (PSF) which is to be
used in the object classification procedure. A specific program in
FOCAS automatically searches for stellar and non-saturated objects for
the construction of the PSF. In this work, we limited that search for
stars to the magnitude interval of B 16$^{\rm m}-19^{\rm m}$.  The
stars selected by FOCAS were visually inspected and after pruning
asymmetric objects we then recomputed the PSF.  The classification of
objects was then performed according to specific rules in FOCAS that we
kept at their default values (see \cite{val82} for a detailed
description). After this process, a visual inspection was carried out
to eliminate catalogued objects that were clearly misclassified by
FOCAS. In general, these objects were overlapping stars and/or galaxies
forming a unique diffuse object mistaken for a galaxy. This type of
contamination is estimated as $\sim$ 10 \% of the whole catalog (see
also \cite{dCAZ94}).

\subsection {Magnitude Calibration and Limiting Magnitude}

The calibration of the magnitude scale was done using CCD photometry
from Postman \& Lauer (1995) (hereafter PL) and Melnick \& Quintana
(1984) (hereafter MQ). PL obtained their data through actual
photometric measurements of the cD galaxy (B-Johnson system) whereas
the data from MQ is a compilation of measurements from the literature
(in different passbands). Based on this, we decided to use the data
from PL to do actual calibration, and to use the MQ data as a
consistency test. Calibration was done by estimating the difference
between the magnitude listed by PL and MQ and the instrumental total
magnitude.  FOCAS calculates the total luminosity by integrating the
flux inside the ``total area'' of the object. This ``total area'' is in
turn determined by fitting the concavities in x,y of the isotope shape
and then adding rings around the object up to 2 times the detection
area (called ``grown area'').  Consequently, the FOCAS instrumental
magnitude beyond the grown radius is constant. In this case PL provides
measurements beyond the radius corresponding to the total area.  For
this reason, we have used the task ``apphot'' in IRAF to measure the
instrumental magnitude within specific apertures.  The internal
apertures were not used due to saturation of the cD galaxy. The
zero-point of the magnitude scale resulted in K = 32.86 $\pm$ 0.16
(using data from PL) and K = 32.92 $\pm$ 0.18 (using data from MQ). The
comparison between the photometry results of PL, MQ and this work is
shown in Figure~\ref{photom} (top), with corresponding residuals
(bottom).

The limiting magnitude of the catalog (B$_{lim} \sim 21$) was first
estimated using the magnitude histogram of all detected objects not
classified by FOCAS as ``noise'' objects, and was defined by the last
bin prior to a significant counting decrease (\cite{pic91}). This
magnitude limit estimate will be reanalysed in the next section.

\subsection {The Final Galaxy Catalog}

The photometric catalog described above was obtained by running FOCAS
under its default classification rules by which the detected objects
are assigned to 5 pre-defined ``resolution'' classes:
n - noise, s - stars, sf - fuzzy stars, g - galaxy, and d - diffuse. 
A further class, ``long'', is
defined for those objects presenting flat and very elongated
geometries.  As discussed in detail by Valdes (1982), the uncertainties
of such a classification are due, for one part, to the natural
uncertainties in the partitioning of the image parameter space into
astronomical classes and, for the other, to the rapid decrease of
the classification probabilities at fainter magnitudes.

In order to circumvent these problems, which are indeed common to most
automated astronomical image classifiers, we used a sample of 325
spectrocopically confirmed galaxies to provide an independent check to
the FOCAS classifier performance. This sample was obtained by
compilating the following redshift catalogs: Bardelli et al. (1993),
Mazure et al. (1997), Metcalfe et al. (1987),  Melnick \& Quintana
(1984), Stein et al. (1996) and Teague et al. (1990). By
cross-correlating it with our photometric catalog we produced a
``matched'' catalog consisting of 322 galaxies. For the remaining 3
galaxies, 2 were not included in the matched catalog because they were
identified as multiple systems not split by FOCAS, and the other one
could not be identified with any detected object. The statistics of
FOCAS classification of these 322 objects is given in Table~\ref{yyy}
(note that no galaxies were classified as ``n'').  As can be seen, the
bright galaxies were correctly classified as ``g'' with a negligible
contamination of the other classes. In contrast, faint galaxies are
increasingly classified as ``sf''. In view of these results, we
conservatively decided to keep in the final catalog only objects
classified as ``g'' and ``sf''.  This final galaxy catalog with 1421
galaxies is available from the AAS CDROM.  We found, using the same
methodology as Picard (1991), that the limiting magnitude of our
catalog of galaxies is about the same as that of the initial
photometric catalog (section 2.3), that is, $B_{lim}\sim$ 21.
 
We have carried out a similar completeness analysis for the subset of
galaxies with measured velocities. We found that it is only moderately
complete in all magnitudes ranges (see Figure~\ref{compl.his}).
Nevertheless, the completeness level per magnitude interval increases
for galaxies systematically closer in projection to the central cD
galaxy. For instance, a $\sim 88$ \% cumulative completeness level up
to the 17.5-18 mag bin is achieved for the galaxies within a 900 arcsec
radius around the cD, whereas a $\sim 95$ \% level up to the 18-18.5
mag bin is found using a 500 arcsec radius. These features are due to
the inhomogeneities of the assembled velocity sample and will be taken
into account in the dynamical analysis.
 
\section {The Velocity Distribution}

We have used the robust estimators discussed by Beers et al. (1990) to
establish the membership of the final catalog of galaxies with measured
velocities (322 galaxies) and to obtain the global kinematical
parameters of A3558. A histogram of all measured velocities for the
matched catalog (inset, Figure~\ref{vel1.ps}) was first used to
eliminate the obvious foreground and background objects (v $<$ 9000 km
s$^{-1}$ and v $>$ 18000 km s$^{-1}$). We then used the ROSTAT code
(\cite{bee90}, \cite{bee91}) to compute the $C_{BI}$ and $S_{BI}$
estimators for the remaining velocities and recursively eliminated the
galaxies that were $3$ biweighted scales from $C_{BI}$ (\cite{tea90}),
which converged to a membership criteria of 12000 km s$^{-1}~ <$ v $<$
18000 km s$^{-1}$.  The final ``velocity'' catalog contains 282
possible member galaxies brighter than $\rm B_{\rm lim} \sim 21$.  Its
velocity histogram is shown in Figure~\ref{vel1.ps}.  For comparison,
the velocity histogram of galaxies within a 900 arcsec radius circle
around the cD galaxy is shown on Figure~\ref{vel2.ps}. As argued in the
previous section, this central subset ($R < 900$ arcsec) is $\sim$ 88\%
complete up to the 18 mag bin and may be taken as a representative
sample of the core of A3558.

At this point we should note that the 1$^{\circ}\times~$1$^{\circ}$
field covered by our photometry also includes, other than A3558 itself,
the poor cluster SC1327-312 (RA = $13^h 30^m 04^s$, DEC = $-31^{\circ}
44^{\prime} 49^{\prime \prime}$), first noted as a X-ray source by
Breen et al. (1994). The velocity histogram of galaxies inside a
circular region of 400 arcsec radius centered at the peak X-ray
emission of SC1327-312 (\cite{bar96}) is displayed in
Figure~\ref{vel3.ps}.  This sample, also defined for galaxies in the
velocity range 12000 km s$^{-1}~ <$ v $<$ 18000 km s$^{-1}$, contains
17 galaxies with measured velocities and is 82\% complete down to 18
mag.

In Table~\ref{tbl-1} the global kinematical parameters of the 3 samples
discussed above are presented. The columns are as follows:  (1), sample;
(2) and (3), $C_{BI}$ and $S_{BI}$ (in km s$^{-1}$) respectively, with
their 90\% confidence intervals; (4), numbers of galaxies used for the
estimate and (5), source reference. We will assume the redshift
corresponding to the central  ($R < 900$ arcsec) region as that
representative for the cluster: $z = 0.0475 \pm 0.0003$.

We have performed several normality tests for the velocity distribution
samples defined above using the ROSTAT code.  The results indicate that
the velocity distribution of A3558, using all 282 probable member
galaxies, is consistent with normality under most of the statistical
tests included in the ROSTAT routine (significance levels for the null
hypothesis greater than 25\%). The scaled tail index, TI = 0.983,
indicates that the sample distribution is close to gaussian, whereas
the asymmetry index, AI = 0.971, suggests that the distribution is
skewed towards higher velocities (for a thorough discussion on these
two indexes, see \cite{bir93}).  Similarly, we find that the ``core''
$R < 900$  arcsec sample (Figure ~\ref{vel2.ps}) is also approximately
gaussian, with TI = 0.942 and AI = 0.996.  The Dip test indicates that
the probability of non-unimodality of the ``core'' sample is very large
($\sim$ 83\%) and this result is in accordance with the presence of a
bimodal core, as reported in detail in the next sections. The
SC1327-312 sample  (see Figure ~\ref{vel3.ps}) displays an asymmetric
distribution, AI = -0.669. We find a TI = 0.768 for this sample, which
indicates that the tails are probably underpopulated, if the parent
distribution is indeed gaussian, and the Dip test indicates that the
probability of  non-unimodality is reasonably high ($\sim$ 70 \%).

In order to discuss the importance of gaps in velocity space, we have
followed the weighted gap analysis presented by Beers et al. (1992).  A
{\it single} gap is considered significant if its gaussian weighted
value is greater than 2.25. Significant gaps found by sampling a
gaussian distribution occur with a frequency less then 3\%,
independently of the sample size. The ``total gap probability'' (TGP)
is another indicator of gap significance: it is the cummulative
probability of finding a {\it number} N of individually significant
gaps anywhere in the distribution. For instance, a low TGP indicates
that, after resampling, all the N original gaps will very unlikely be
found anywhere in the distribution, that is, the N gaps will be
collectivelly significant.  We use the ``stripe'' density diagram to
show the location of each significant gap (\cite{bee92}).  Each galaxy
in velocity space is indicated by a unique vertical line, and 
significant gaps are indicated as arrows.

In the 282 galaxy sample velocity distribution (see upper diagram of
Figure~\ref{vel1.ps}), we find that the probability of each individual
gap is 0.03, except for one gap at $\sim$ 13700 km s$^{-1}$, which is
0.002. Nevertheless, the ``total gap probability'' indicates that the
probability that all these gaps occur anywhere in the distribution is
greater than $\sim$ 40\%.  The ``core'' sample has several gaps.  The
position, size, weighted size and probability of each gap for this
sample are:  13570 km s$^{-1}$ ($\Delta$v = 79 km s$^{-1}$, $z_{*}$ =
2.297, p = 0.03),  15501 km s$^{-1}$ ($\Delta$v = 172 km s$^{-1}$,
$z_{*}$ = 2.457, p = 0.03), and 14505 km s$^{-1}$ ($\Delta$v = 73 km
s$^{-1}$, $z_{*}$ = 2.470, p = 0.03), 14640 km s$^{-1}$ ($\Delta$v = 83
km s$^{-1}$, $z_{*}$ = 2.580, p = 0.014), 14425 km s$^{-1}$ ($\Delta$v
= 84 km s$^{-1}$, $z_{*}$ = 2.658, p = 0.014), and finally, 14763 km
s$^{-1}$ ($\Delta$v = 105 km s$^{-1}$, $z_{*}$ = 2.803, p = 0.006).
The ``total gap probability'' is  $\sim$ 21 \% in this case: the
occurence of all these gaps anywhere in the distribution after
resampling is relatively small.  The SC1327-312 sample shows no
significant gaps.

The presence of gaps indicates that the distribution probably has several
``clumps'' in the velocity space, but their overlapping makes
separation impossible without positional information.  A complete
substructure analysis, using velocity/position data, is presented in
the following section.

\section {Substructure Analysis}

\subsection{Introduction}

Clusters of galaxies can be morphologically arranged into an
one-\-di\-men\-sional sequence (from irregular to regular) that
suggests a direction of dynamical evolution (from young to dynamically
evolved systems). A virialized cluster is also expected to have a
gaussian line-of-sight velocity profile, whereas a young system should
present significant deviations from a gaussian profile in its velocity
distribution. Although separate positional and velocity investigations
can yield general clues about the global dynamical status of a cluster,
only 3-dimensional $(x,y,v)$ diagnostics can reduce the possiblity of
misinterpretation, since only combined correlations between position
and velocity may define physical substructures, suggesting a
non-equilibrium state of the cluster.  Recent multi-fiber spectroscopy
has allowed the construction of large scale redshift surveys
(\cite{zab93}, \cite{maz97}) which, added to positional and photometric
data, contributes to a reliable approach to the problem of substructure
in clusters of galaxies (\cite{bir94}).  Indeed, recent optical and
X-ray obervations (\cite{wes94b} and references therein) suggest that
up to $75$\% of clusters show substructures in their morphologies.

The presence of substructures can also be analysed via global
statistical tests which yield a substructure significance index for the
data. These tests are especially useful for a general and direct
comparative analysis of a large sample of clusters (see, e.g.
\cite{bir94}), but they do not pinpoint the location, extension and
dynamical nature of the substructures present in the data. For this
reason, we shall adopt mapping techiques which  are quite useful for
this purpose.

\subsection {Substructure Mapping and Dynamical Analysis}

We use the Adaptive Kernel technique (AK, \cite{sil86}) with a
generalised Epanechnikov kernel in order to map the projected density
of galaxies and also the local average velocities and the local
velocity dispersion of galaxies (\cite{biv96}). Following Biviano et
al. (1996), we have obtained significance maps for these quantities by
taking the average of 1000 bootstraps of the original cluster and
subtracting the corresponding $3\sigma$ maps from this average.
Structures that can ``survive'' this subtraction are statistically
significant features at the $3\sigma$ significance level.

We have divided our field into three regions, namely: the {\it core
region} defined by a square field of 1000 arcsec side centered on the
cD galaxy, and encompassing most of the x-ray emission of the cluster.
In this region the velocity sample is 89\% complete down to B $\sim$
19; an {\it intermediate region} defined by a square field of 1800 arcsec
side centered on the cD galaxy for which the velocity sample is 88\%
complete down to B $\sim$ 18; and {\it the whole
1$^{\circ}\times~$1$^{\circ}$ field}, for which the velocity sample is
not complete at any limiting magnitude.  All mappings have grids with 1
arcmin ($\simeq 0.05 h_{75}^{-1}Mpc$) spacing. These AK maps are
presented in Figs.  6-13, where in all figures the original cluster map
is in the upper panel and the $3\sigma$ confidence map in the lower
panel.

In order to analyse the substructures present in these figures, we
use the Wavelet Transform technique (WT, \cite{gro87},
\cite{sle90}) to generate the subcatalogs of galaxies for each
substructure of interest (namely, the central substructures and the
poor cluster SC1327-312). These subcatalogs were constructed by
selecting galaxies inside a $2s$ radius around density peaks, where $s$
is the wavelet scale which best characterizes the analysed structure
(see \cite{esc94} and \cite{sle90} for details).
 
The core region (Figs. \ref{dxyc}-\ref{dsvc}) features a significant
bimodal structure (denoted by A and B in all maps). This central region
is thoroughly analysed in the following subsections.  The apparent
central bimodal core is also found in the intermediate region map
(Figure \ref{dxyi}), where several significant substructures (at
$3\sigma$ significance level) are also present.  Statistically
significant average velocity and velocity dispersion structures can
also be noted (Figures \ref{dvmi} and \ref{dsvi}, respectively). 
The whole field around A3558 is clearly marked with several
substructures (Figure \ref{dxyw}). The poor cluster SC1327-312
towards the southeast, known {\it a priori} to be partially
represented on our field, is also clearly detected.
Unfortunately the incompleteness of the velocity catalog for the
whole region does not allow us to assess the reality of most these
substructures. However, as we argue below, their abundance and
projected distribution suggests that the infall region of A3558 ($>
1 ~$ Mpc) may be composed of several groups of galaxies in the
process of merging. 

Indeed, we qualitatively note a ``preferential alignment'' at the
$\sim$45$^{\circ}$ position angle of a diverse set of features in this
cluster, ranging from small to large scales. An application of the
Lee-statistics (\cite{fit88}) to this cluster also indicates that the
$\sim$45$^{\circ}$ direction has a greater probability of bimodality.
This alignment coincides with (a) the major axis PA of the dominant
galaxy, (b) the major axis PA of the isocontours of the central core
(using AK/WT maps as well as a X-ray countour map), (c) a marginal
velocity gradient direction across the central bimodal substructures,
and (d) the direction of alignment of major subclumps, namely the
bimodal core and SC1327-312.  This alignment seems to persist even
beyond the analysed field. A qualitative inspection of an isopleth map
of a $\sim 32^{\circ} \times 32^{\circ}$ region around A3558 (see
Figure 2 of \cite{ray91}), also indicates an alignment of all major
clusters in the Hydra-Centaurus region spanning from position angles of
$\sim 40^{\circ}$ to $\sim 60^{\circ}$. Considering the physical
dimensions involved, it is quite surprising that the average alignment
of major clusters in the Shapley Concentration is consistent with the
direction of alignment of several substructures within A3558 and with
the major axis position angle of its dominant galaxy. We find that
these small to large scale ``coincidental'' associations (namely, cD
major axis match to the general clustering alignment) can be taken as
an observational evidence of an anisotropic merger scenario as for
instance the one proposed by West (1994a).

\subsubsection {The complexity of the central core of A3558}

Figure \ref{dxyc} (upper panel) presents the projected density contour
map of the very central core of A3558. It shows a clear bimodal
structure which is very proeminent on the $3\sigma$ confidence map, as
can be seen in the lower panel of this figure. The two central clumps
are labeled ``A'' (RA = $13^h 27^m 49^s$, DEC = $-31^{\circ}
28^{\prime} 59^{\prime \prime}$) and ``B'' (RA =  $13^h 28^m 13^s$, DEC
= $-31^{\circ} 33^{\prime} 10^{\prime \prime}$) in Table \ref{beta}.
The average velocities  of these central substructures are also
significantly distinct, as can be seen in Figure \ref{dvmc} and also by
an inspection of the $C_{BI}$ values displayed in Table \ref{beta} (the
difference of their average velocities are out their 90
central location).  The isodispersion map (Figure \ref{dsvc}) indicates
that the average velocity dispersion of substructure A differs from
that of the central region, although not very significantly.  This is
also in accordance with the $S_{BI}$ values listed in Table
\ref{beta}.

Figure \ref{dxyc_faint} displays the projected density contour map of
the faint galaxy subsample of the core region ($19 < B < 21$).  The
distribution of faint galaxies is much smoother than that of the
brighter galaxies, having peak densities which attain no more than half
of those of the bright subsample. Both substructures A and B are still
present in this map, but much less prominently.  There are also hints
of new significant clumps which are mapped only by faint galaxies.
Unfortunately, this faint central subsample contains only 5 galaxies
with measured velocities (all of them in the range 12000 $<$ v $<$
18000 km s$^{-1}$) precluding a more detailed analysis about the
possible existence of background clusters.

The above discussion stresses the apparent complexity of the central
core of A3558. This is to be contrasted to the featureless X-ray
emission, Figure \ref{xray}, which is centered on the cD galaxy with a
smooth extension towards the location of substructure B (see also the
detailed study by \cite{bar96}). This suggests that the gas is driven
by a smooth gravitational potential which does not seem to be mapped by
the bright galaxy component of the cluster. We are thus led to
conjecture that the projected core of A3558 is made of 3 dynamically
distinct structures: the main virialized core of the cluster, which is
traced by the X-ray gas and maybe also by the faint galaxy component;
and the galaxy clumps A and B, described before.

\subsubsection {Solving the cD offset}

To support the picture described above, we have verified that the
velocity histogram of substructure A is apparently bimodal (see Figure
\ref{histA.ps}). Indeed, as noticed previously, the asymmetry index of
the velocity distribution of the $R < 900$ arcsec sample is 0.996. This
suggests that there is a high velocity population excess, and this may
be partially due to substructure A. The statistics of substructure A
velocity distribution indicates a scaled tail index at 0.886 and no
significant gaps where found.  We notice that one of the two peaks of
the velocity distribution of substructure A is centered very nearly at
the velocity of the cD. The Dip test applied to this sample indicates
that the probability of non-unimodality is $\sim$ 30 \%. Both peaks
also seems to have a reasonably narrow dispersion, suggesting that
clump A galaxies may be separated into two subsamples:  galaxies with v
$<$ 14750 km s$^{-1}$ belong to the core sample defined above, and
galaxies with v $>$ 14750 km s$^{-1}$ form the actual background group
A.  This group is denoted in Table \ref{beta} as A$^{\prime}$.
 
Figure \ref{new_core} shows the isodensity map which results from
removing the A$^{\prime}$ galaxies from the core region (this new
sample is denoted in Table \ref{beta} as ``core - A$^{\prime}$''). 
There is no sign of any clumping at the location where
substructure A was previously detected. The average velocity of this
core sample is almost exactly the same as that of the cD galaxy
(14030 $\pm$ 42 km s$^{-1}$), that is, the velocity offset
of the cD is contained within the 90 \% interval on central location
in velocity space and is no longer significant (see the confidence
limits on Table \ref{beta}.

\subsubsection {The $\beta$-problem revisited}

The velocity dispersion for the new core sample discussed above is much
smaller than the value generally quoted for A3558 (see Table
\ref{beta}) which, for instance, led ardelli et al. (1996) to postulate a
``strong $\beta$-problem'' for A3558. Assuming the new value given in
Table \ref{beta}, we find $\beta_{spec}$ = 1.13 (0.93,1.31), to be
compared with $\beta_{fit}$ = 0.611 estimated by Bardelli et al. (1996)
from fitting the X-ray brightness to an isothermal model. A
$\beta$-problem still remains, but can no longer be considered
``extreme''(the $\beta_{spec}$ value found by \cite{bar96} for A3558 is
1.79).

Nevertheless, the above results may be improved by further removing
substructure B galaxies from the core sample.  This removal may be
justified if we interpret the X-ray extension towards B  as due to the
gas response to a local fluctuation of the gravitational potential due
to B. This interpretation is corroborated by the WT analysis of the
X-ray emission of A3558 given by Slezak et al. (1994), which shows
evidence for a X-ray counterpart for substructure B. Since the
dynamical time of collision for the gas is much shorter than for the
galaxies, we can assume that the infall of subclump B galaxies is still
a recent event and has not strongly disturbed the dynamics of the
core.  The kinematical parameters for the sample obtained by removing
substructure B from the core defined before are given in Table
\ref{beta}. Notice that the difference between the cD velocity and the
mean velocity of the core is further reduced for this sample. We now find
$\beta_{spec}$ = 0.89 (0.66,1.22) which is much closer to $\beta_{fit}$
than the previous calculated value.  Moreover, if we take into
account the more reliable ASCA temperatures for A3558 given by
Markevitch, Sarazin, \& Henriksen (1996), we will find that
$\beta_{spec}$ = 0.70 (0.55,0.85), a value which coincides almost
exactly with $\beta_{fit}$ = 0.611 estimated by Bardelli et al. Notice
that using the global core velocity dispersion (i.e, without removing
substructure B), together with ASCA temperatures, gives $\beta_{spec}$ =
1.09 (0.89,1.29), to be compared to the value quoted by Bardelli et al. of
1.79, using the more uncertain ROSAT temperatures (\cite{mar96}).

According to Bahcall \& Lubin (1993), a solution for the
$\beta$-problem arises naturally if an empiric spatial $r^{-2.4}$ law,
fitted for the cluster outer region ($> 1.5 Mpc$), is used instead of a
King law. Since we found that the $\beta$-discrepancy for A3558 could
also be naturally diminished by using a better estimation of the
velocity dispersion, we analysed in detail the density profile to check
the above hypothesis.  We compared the projected density profile of
galaxies in the central region to the X-ray brightness profile. This
profile, given by Bardelli et al. (1996), is an elliptical King law fit
to the overall cluster emission.  By taking the axial ratio and
position angle derived from their fit, we find that the projected
central density profile of galaxies could also be very well fitted by a
King law. This is in agreement with the hypothesis that our selected
sample represents the relaxed core component, with the galaxies in
equilibrium with the emitting gas.  However, the galaxy sample selected
from the whole region of our plate could only be fitted by the sum of 2
King profiles, one of which clearly represents the core component
whereas the other, with have a very large core radius, probably
represents a background component. At the wings of the projected galaxy
distribution, the resulting combined profile tends towards a $\sim
r^{-1.5}$, very similar to the one proposed by Bahcall \& Lubin.

\subsection {Mass Estimates}

We derive mass estimates for A3558 under the hypothesis that (1) the
system is self-gravitating (we refer to this mass estimate as
$M_{selfgrav}$; \cite{hei85}) and (2) the galaxies orbit around a
common central potential ($M_{sat}$; \cite{bah81}).  The mass estimates
are calculated for increasing circular regions centered on the cD,
after removing galaxies belonging to substructure A$^{\prime}$.  We have
also calculated the mass of the poor cluster SC1327-312. The resulting
values are listed in Table \ref{tbl-3}.  Columns are: (1) label; (2)
aperture radius in Mpc centered on the cD; (3) number of galaxies
defining the region; (4) and (5) $C_{BI}$ and $S_{BI}$ in km s$^{-1}$,
respectively together with their 90\% confidence limits; (6) mass in
units of $10^{14} ~h^{-1}_{75} ~M_{\odot}$, according to \cite{hei85}
and estimated error; (7) mass in units of $10^{14} ~h^{-1}_{75}
~M_{\odot}$, according to \cite{bah81} and estimated error; and (8)
$M_{sat}$/L ratio in units of $10^{3}~ M_{\odot}/L _{\odot}$,  the
total luminosity being computed  down to B = 19.

Figure \ref{mass.ps} shows the comparison of our results to others
found in the literature.  Our results, based on the self-gravitating
hypothesis, are compatible (although systematically higher) with
previous dynamical estimates found by Biviano et al. (1993), Metcalfe
et al. (1987) and Metcalfe et al. (1994), based on different samples.
Although systematically higher by a factor $\sim$ 2, our results based
on the ``satellite'' hypothesis provide lower values for the  mass
estimates, which are more compatible with the mass estimates provided
by Bardelli et al.  (1996), based on X-ray data.  This is consistent
with the fact that most of the cluster mass is not bound to the visible
galaxies.   The discrepancy between our mass estimates  compared to the
X-ray total mass estimates may be partially explained by the presence
of substructures which were not well taken in account in our
calculations (\cite{bir94}). Indeed, only substructure A$^{\prime}$ could
be reasonably isolated in velocity space. We note that, within errors,
the M/L ratio is approximately constant with radius distance, averaging
to about 250 $M_{\odot}/L_{\odot}$ (the poor cluster SC1327-312 also
has a M/L ratio consistent with that of A3558). This suggests that
galaxies follow the same distribution as the total cluster mass, as
already noted by Bardelli et al. (1996).

\section{Discussion}

In this paper, we examine the cluster of galaxies A3558, focusing on
substructure analysis, using position and velocity data. We find that a
simple analysis of the velocity distribution is not sufficient to
resolve the cluster into multiple components. Only a 3-D analysis can
establish the nature of these subsystems.

The AK isodensity, isovelocity and isodispersion maps and the WT analysis 
all indicate that the region defining A3558 is actually composed of a
collection of several groups of galaxies, suggesting that
A3558 is a dynamically complex, young cluster of galaxies. This
conclusion is consistent with recent results (\cite{bar96},
\cite{mar96}) using independent (namely, X-ray) data.  

We also find a ``preferential alignment'' at the $\sim$45$^{\circ}$
position angle of a diverse set of features in this cluster, ranging
from small to large scales. In addition to this result, the number
statistics of the bright X-ray clusters in the Hydra-Centaurus region
also demonstrates the presence of a linear feature which is not a
chance aggregation of clusters (\cite{ray91}).  As Shapley himself
noted, the supercluster which bears his name is notable due to ``its
great linear dimension, the numerous population, and distinctly
elongated form'' (\cite {sha30}). Although standing only as a
qualitative analysis, our results can be taken as observational support
of an anisotropic merger scenario (\cite{wes94a}; see also
\cite{rhe92}, \cite{pli94}).  Other observational support for such
scenario comes from an
analysis of correlations between the spatial distribution of
substructures and the surrounding matter found in a sample of 7 X-ray
clusters with at least three major subclumps  (\cite{wes95}).  Other
similar examples of substructure/large scale distribution alignments
are: the Coma cluster (\cite{wes95}), A426 (\cite{sch92}) and AWM7
(\cite{ste95}).

Our analysis of the cluster core clearly indicates that the cD velocity
offset is no longer observed. This is interesting because it is
generally supposed that dominant cD galaxies in clusters are formed out
of debris of disrupted galaxies orbiting near the center-of-mass (CM)
of the clusters and/or by accretion of intracluster gas from a cooling
flow, so their velocities should reflect that of the cluster CM. Also,
this result is compatible with the analysis of \cite{geh91}, where the
authors suggested that a large fraction of clusters in their data
sample that presented significant cD offsets could be examples of
clusters with real substructure. Our results are also compatible with
the study of \cite{bir94} on the correlation between the cD offset and
the degree of substructuring.

Concerning the $\beta$-problem, although the incompleteness in velocity
data of our catalog at regions greater than 1.5 Mpc forbids a direct
check of Bahcall \& Lubin (1993) hypothesis, a background component
could be responsible, at least partially, for the observed $r^{-2.4}$
law.  However, only a complete velocity sample for the outer cluster
region can establish whether the $r^{-2.4}$ law represents a real trend
in cluster density profiles or not. Such analysis would be extremely
important for our understanding of cluster formation.

It is interesting to note that our results imply that:  (1) the
intracluster gas and the galaxies are in approximate thermodynamic
equilibrium ($\beta_{spec}$ = $\beta_{fit}$); (2) the intracluster gas
is more spread in phase-space than the galaxies ($\beta_{spec}~ <~ 1$).
This implies no $\beta$-discrepancy for A3558, but a mechanism to
explain why $\beta_{spec}~ <~ 1$ is necessary. An early satellite
merger with A3558 is suggestive, since the ellipsoidal external
x-ray isophotes of the cluster could be interpreted as a merger
signature.  Finally, our results indicate that a superestimation of the
velocity dispersion due to substructure contamination is clearly an
important factor contributing to the arisal of the $\beta$-discrepancy
in clusters.

\acknowledgments

We thank Timothy Beers for providing the ROSTAT code and the first
version of the Adaptive Kernel program. His valuable referee comments
were also very helpfull for improving the final version of this
manuscript and we thank him for that. We also thank Andrea Biviano for
discussions about the bootstrapping techniques used in this work.  One
of us (C.C.D.) acknowledges a fellowship from CAPES.  This work has
been partly supported by a CNPq-CNRS grant.

\clearpage
\figcaption{Comparison between the photometry results of PL, MQ and
this work, with corresponding residuals (where $\Delta B$ is the
difference between ours and PL, MQ results).  \label {photom}}

\figcaption{Magnitude histograms of the photometric catalog (solid
line), including all detected objects classified as ``g'' and``sf'',
and the velocity matched catalog with 322 galaxies (dotted line).  The
magnitude limit based on Picard's methodology (1991) is marked with an
arrow.  \label {compl.his}}

\figcaption{{\it Inset:} Velocity histogram of the matched catalog with
322 objects. {\it Figure:} Velocity histogram for the final 3
$S_{BI}$-clipped matched catalog with 282 objects.  {\it Upper
diagram:} ``Stripe'' density plot for the velocity distribution; gaps
are indicated as arrows. \label {vel1.ps}}

\figcaption{Velocity histogram for the $R < 900$ arcsec sample. {\it
Upper diagram:} ``Stripe'' density plot for the velocity distribution;
gaps are indicated as arrows. \label {vel2.ps}}

\figcaption{Velocity histogram for the SC1327-312 sample. 
{\it Upper diagram:} ``Stripe'' density
plot for the velocity distribution; gaps are indicated as arrows.
\label {vel3.ps}}

\figcaption{Isodensity contour map using the Adaptive Kernel technique
(top) and its $-3 \sigma$ significance level map (bottom) for the {\it
intermediate region} centered at the cD galaxy (represented by a solid
square). North is up and east to the left; both axis have ticks equally
spaced by 300 arcsecs.{\it Top}: contoured from 0 to 2 $\times 10 ^{-4}$ gal
deg$^{-2}$, in intervals of 2 $\times 10^{-5}$. {\it Bottom}: contoured from
0 to 9.6 $\times 10 ^{-5}$ gal deg$^{-2}$, in intervals of 6 $\times
10^{-6}$. \label {dxyi}}

\figcaption{Same as Figure 6 for the local average velocity contour map.
{\it Top}: contoured from 13100 to 15600 km s$^{-1}$, in intervals of
200 km s$^{-1}$.  {\it Bottom}:  contoured from 10000 to 14200
km s$^{-1}$, in intervals of 600 km s$^{-1}$.
 \label {dvmi}}

\figcaption{Same as Figure 6 for the local velocity dispersion contour
map.  {\it Top}: contoured from 0 to 1800 km s$^{-1}$, in intervals of
200 km s$^{-1}$.  {\it Bottom}: contoured from 0 to 1000 km s$^{-1}$,
in intervals of 200 km s$^{-1}$.
 \label {dsvi}}

\figcaption{Isodensity contour map using the Adaptive Kernel technique
(top) and its $-3 \sigma$ significance level map (bottom) for the whole field
of A3558 ($B_{lim}=19$). The position of the cD galaxy is represented by a
solid square. North is up and east to the left; both axis have ticks equally
spaced by 500 arcsecs. {\it Top}: contoured from 0 to 4 $\times 10 ^{-4}$ gal
deg$^{-2}$, in intervals of 2 $\times 10^{-5}$.  {\it Bottom}: contoured from
0 to 2.6 $\times 10 ^{-4}$ gal deg$^{-2}$, in intervals of 1.3 $\times
10^{-5}$. \label {dxyw}}

\figcaption{Isodensity contour map using the Adaptive Kernel technique
(top) and its $-3 \sigma$ significance level map (bottom) for the central
{\it core region} centered at the cD galaxy (represented by a solid square).
North is up and east to the left; both axis have ticks equally spaced by 200
arcsecs.  {\it Top}: contoured from 0 to 4 $\times 10 ^{-4}$ gal deg$^{-2}$,
in intervals of 2 $\times 10^{-5}$.  {\it Bottom}: contoured from 0 to 1.4
$\times 10 ^{-4}$ gal deg$^{-2}$, in intervals of 2 $\times 10^{-5}$. \label
{dxyc}}

\figcaption{Same as Figure 10 for the local average velocity contour map.
{\it Top}: contoured from 13100 to 15700 km s$^{-1}$ , in intervals of
200 km s$^{-1}$.  {\it Bottom}: contoured from 10000 to 14500
km s$^{-1}$ , in intervals of 500 km s$^{-1}$.
 \label {dvmc}}

\figcaption{Same as Figure 10 for the local velocity dispersion contour
map. {\it Top}: contoured from 100 to 1300 km s$^{-1}$, in intervals of
200 km s$^{-1}$.  {\it Bottom}: contoured from 0 to 675 km s$^{-1}$, in
intervals of 75 km s$^{-1}$.
 \label {dsvc}}

\figcaption{As in Figure 10, we plot the isodensity contour map for the
{\it central core}  sample of faint galaxies $19 < B < 21$.  {\it Top}:
contoured from 0 to 3.2 $\times 10 ^{-4}$ gal deg$^{-2}$, in intervals
of 2 $\times 10^{-5}$.  {\it Bottom}: contoured from 0 to 1.7 $\times
10 ^{-4}$ galaxies per square degree, in intervals of 1 $\times
10^{-5}$.\label{dxyc_faint}}

\figcaption{X-ray contour map of the central {\it core region} ({\it top}) 
and the whole field ({\it bottom}). Both maps are centered in the cD galaxy
(solid square) and may be directly compared to the isodensity maps of 
Figures 10 and 9 respectively.  \label {xray}}

\figcaption{Velocity histogram for the A sample. {\it Upper diagram:}
``Stripe'' density plot for the velocity distribution.  \label
{histA.ps}}

\figcaption{Isodensity contour map for the central {\it core region} with
substructure A removed as described in the text. Contour levels are the
same as in Fig 10. \label {new_core}}

\figcaption{Mass estimates (in units of $10^{14}~ h^{-1}_{75}
~M_{\odot}$) within given aperture radii (in Mpc) centered around the
cD galaxy, accordingly to several authors. \label {mass.ps}}

\clearpage

\setcounter{table}{0}
\begin{table}
\caption {\label{yyy}}
\end{table}

\begin{table}
\caption {\label{tbl-1}}
\end{table}

\begin{table}
\caption {\label{beta}}
\end{table}

\begin{table}
\caption {\label{tbl-3}}
\end{table}

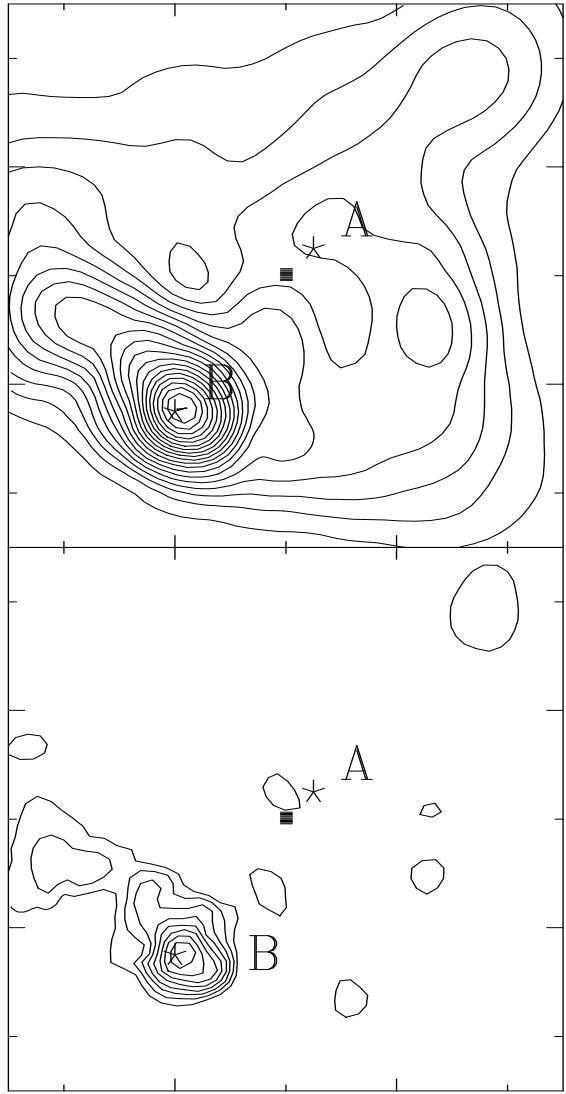
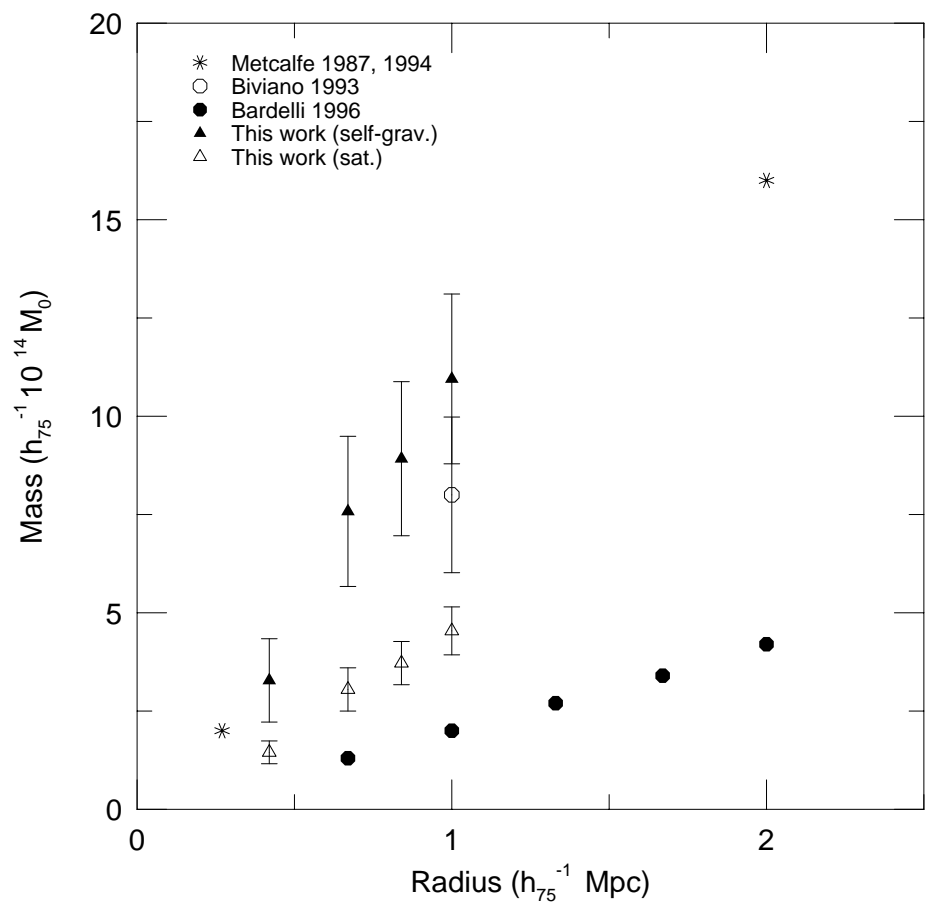
\end{document}